\documentclass[prb,twocolumn,superscriptaddress,showpacs]{revtex4}
\usepackage{graphicx}
\usepackage{amsmath}
\usepackage{color}

\newcommand{\Tr}{\mathrm{Tr}}

\begin{document}

\title{Determinant quantum Monte Carlo study of exciton condensation in the bilayer Hubbard model}
\author{Louk Rademaker}
\email{rademaker@lorentz.leidenuniv.nl}
\affiliation{Institute-Lorentz for Theoretical Physics, Leiden University, PO Box 9506, NL-2300 RA Leiden, The Netherlands}
\author{Steve Johnston}
\affiliation{Department of Physics and Astronomy, University of British Columbia, Vancouver, British Columbia, Canada V6T 1Z1}
\affiliation{Quantum Matter Institute, University of British Columbia, Vancouver, British Columbia, Canada V6T 1Z4}
\author{Jan Zaanen}
\affiliation{Institute-Lorentz for Theoretical Physics, Leiden University, PO Box 9506, NL-2300 RA Leiden, The Netherlands}
\author{Jeroen van den Brink}
\affiliation{Institute for Theoretical Solid State Physics, IFW Dresden, 01171 Dresden, Germany}
\affiliation{Department of Physics, TU Dresden, D-01062 Dresden, Germany}
\date{\today}

\begin{abstract}
We studied the possibility of exciton condensation in a strongly correlated bilayer extended Hubbard model using Determinant Quantum Monte Carlo. To model both the onsite repulsion $U$ and the interlayer interaction $V$ we introduced a novel update scheme extending the standard Sherman-Morrison update. We observe that the sign problem increases dramatically with the inclusion of the interlayer interaction $V$, which prohibits at this stage a unequivocal conclusion regarding the presence of exciton condensation. However, enhancement of the interlayer tunneling results suggest that the strongest exciton condensation tendency lies around $10-20\%$ $p$/$n$-doping. Magnetic properties and conductivity turn out to be relatively independent of the interlayer interaction.
\end{abstract}


\maketitle
\section{Introduction}

The standard Bardeen-Cooper-Schrieffer theory of superconductivity can be straightforwardly extended to describe condensation of electron-hole pairs, known as excitons\cite{Blatt:1962p4000,Keldysh:1968p4790,Moskalenko:2000p4767}. Though the electron-hole Coulomb interaction is naturally quite large, exciton annihilation will suppress the formation of a condensate. To counter the electron-hole recombination problems one could spatially separate the electrons and holes into different layers\cite{Shevchenko:1976p4950,Lozovik:1976p4951}. Recently such \emph{bilayer exciton condensation} has been experimentally verified in quantum Hall bilayers\cite{Eisenstein:2004p4770} and without an externally applied magnetic field in electrically gated, optically pumped semiconductor quantum wells\cite{High:2012p5349}.

The progress in semiconductor quantum wells lead to several theoretical proposals for different candidate materials for exciton condensation, amongst them topological insulators\cite{Seradjeh:2009p4980} and double layer graphene\cite{Lozovik:2008p4877,Zhang:2008p4895,Dillenschneider:2008p4896,Min:2008p4795,Kharitonov:2008p5044}. In this paper we consider bilayer strongly correlated electron systems, such as realized in the multilayer high-temperature superconducting cuprates\cite{Imada:1998p2790,Lee:2006p1688}. Cuprates are quasi-two-dimensional and can be chemically doped with electrons or holes, and it is therefore experimentally feasible to construct heterostructures of differently doped cuprates\cite{Mao:1993p5733,Hoek13}. The strong interactions would effectively reduce the electronic kinetic energy, which favors exciton binding. The physics of correlated $p$/$n$ bilayers is, however, extremely nontrivial. Under the assumption of strong exciton binding energies it has been suggested that exciton condensation in correlated bilayers leads to interesting exciton-spin cooperative effects, reflected in an enhanced triplon bandwith.\cite{Rademaker:2013p5631} 

\begin{figure}
	\includegraphics[width=\columnwidth]{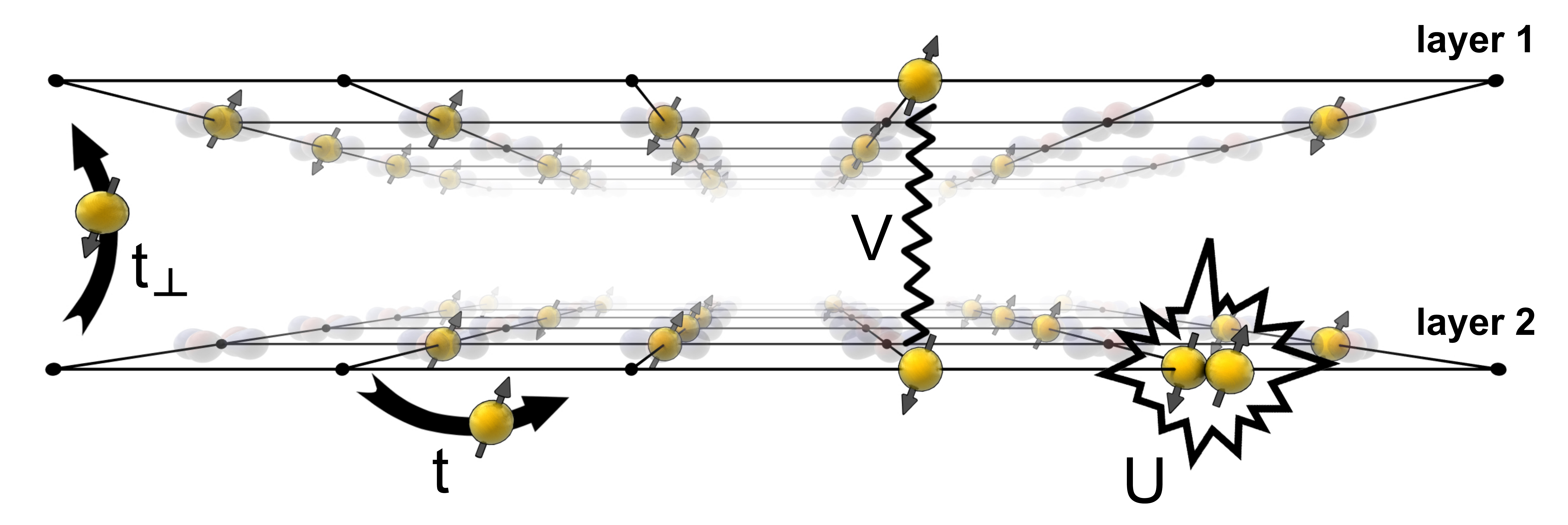}
	\caption{The extended bilayer Hubbard model, Eqns. (\ref{BilayerBeginEq})-(\ref{HVCh3}), describes two layers with in-plane hopping $t$ and interlayer $t_\perp$, onsite repulsion $U$ and interlayer Coulomb interaction $V$. In this paper we study this model using DQMC.}
	\label{BilayerHubbardFig}
\end{figure}

In this context, the major challenge is to study exciton condensation in correlated bilayers when the kinetic energy, onsite repulsion, and interlayer interaction are all treated on the same level. Earlier studies of correlated bilayers are limited to either the weak-coupling regime\cite{Millis:2010p5231} or the strong coupling $t$-$J$ model\cite{Ribeiro:2006p5232,Rademaker:2013p5631}. Numerical simulations considered the bilayer Hubbard model without interlayer interactions\cite{Bouadim:2008p3331} or without onsite repulsion\cite{Kaneko:2013p5630}. Here, we consider both interactions and use the Determinant Quantum Monte Carlo method (DQMC)\cite{Blankenbecler:1981p3063,White:1989p3778,White:1989p4002} to study the extended bilayer Hubbard model as shown in Fig. \ref{BilayerHubbardFig}. In order to include both in-plane and interlayer interactions we developed a novel update scheme as an extension of the usual Sherman-Morrison update. These `Woodbury updates', described in section \ref{SecMeth}, can be used in other models with more than one type of Hubbard-Stratonovich field.

\begin{figure}
	\includegraphics[width=\columnwidth]{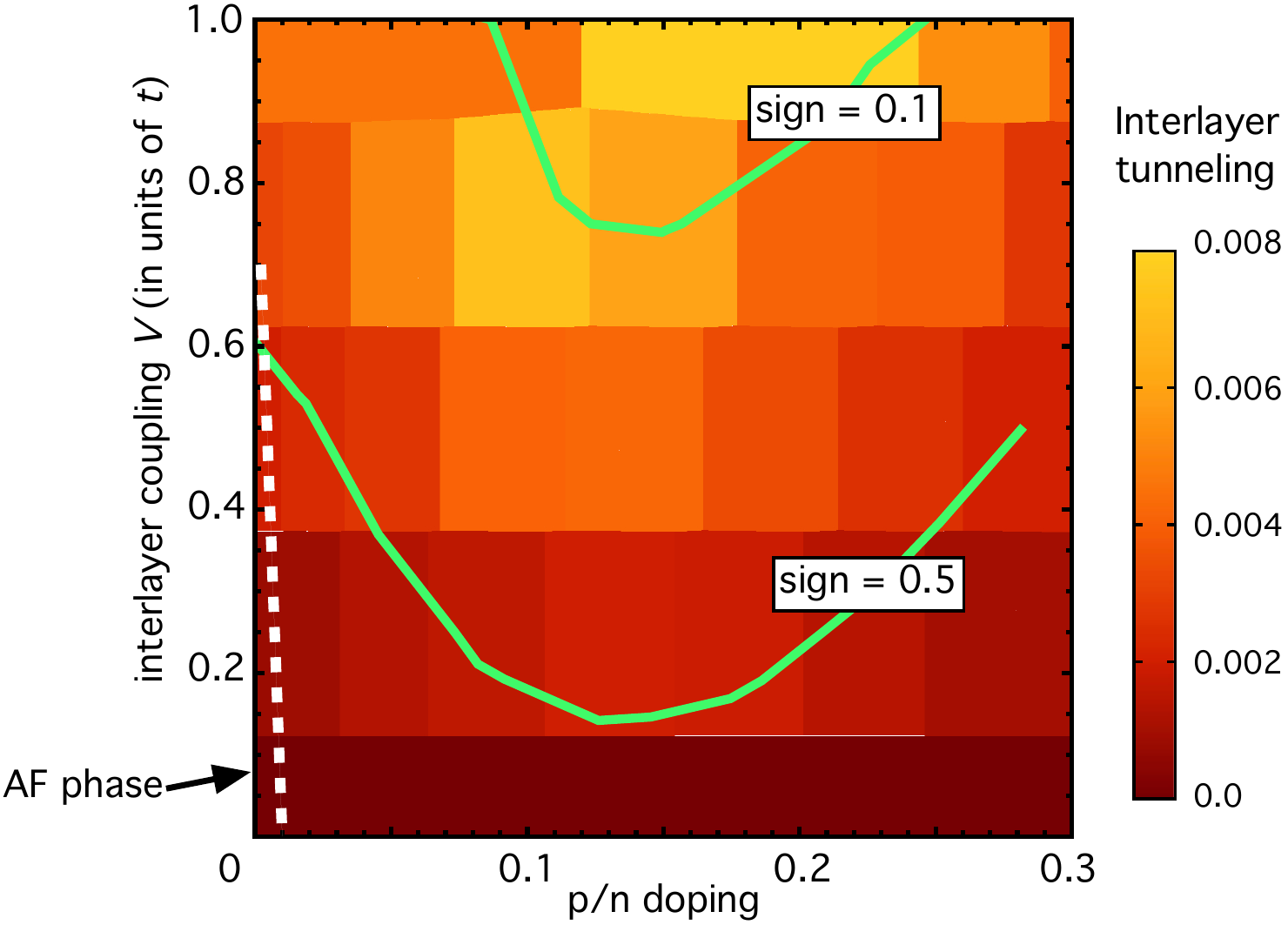}
	\caption{Combining measurements of the interlayer tunneling, the average sign and antiferromagnetic correlations we can construct a tentative phase diagram of the Hubbard biayer, as a function of the interlayer interaction $V$ and $p$/$n$-doping at $T=0.221t$ and $U=4t$.
	The average sign is shown by the solid green lines, indicating the contour lines where the sign is 0.5 or 0.1. The inclusion of $V$ leads to a drastic reduction of the average sign.
	The antiferromagnetic order, where the phase boundary is shown by a dashed white line, is relatively independent of $V$.
	The interlayer tunneling, a signature of exciton formation, increases with $V$ and reaches it maximum around $10-20\%$ $p$/$n$-doping. This figure is a combination of the data shown in Figs. \ref{Fig3-3}, \ref{Fig4-6} and \ref{Fig5-1}.}
	\label{TentativePD}
\end{figure}

To maximize the possibility of exciton condensation, we consider double layered systems where one layer is $n$-doped and the other layer is $p$-doped, with respect to half-filling ($\langle n \rangle = 1$ in each layer). One would expect that with the formation of bosonic excitons the fermionic sign problem would be reduced, thus allowing for numerical control in physically interesting parameter regimes. However, as we show in section \ref{SecFermi} and illustrate in Fig. \ref{TentativePD}, the sign problem is actually enhanced dramatically with the inclusion of an interlayer coupling $V$. Even at moderate temperatures ($T=0.221t$, with $t$ the in-plane nearest-neighbor hopping) and intermediate couplings ($V=0.75t$), the average sign drops to about $\sim 0.1$ around $15\%$ $p$/$n$-doping. Therefore, within the present DQMC approach, one cannot conclude unequivocally whether exciton condensation in $p$/$n$-doped correlated bilayers is possible. We have, however, strong indications that the tendency towards exciton condensation is largest around $10-20\%$ $p$/$n$-doping, as we discuss in section \ref{SecEC}. On the other hand, the magnetic properties of the correlated bilayer turn out to be remarkably independent of the interlayer coupling, as is discussed in section \ref{SecMag} and also illustrated in Fig. \ref{TentativePD}. We conclude this paper with an outlook regarding excitonic physics in correlated bilayers.

\section{Methods}
\label{SecMeth}

\subsection{The Bilayer Hubbard Model}

We begin by introducing the bilayer Hubbard model with interlayer couplings, and discussing the DQMC algorithm used to analyze this model. We consider a bilayer system of two square lattices as shown in Fig. \ref{BilayerHubbardFig}. Each lattice site is denoted by $i \ell$ where $\ell=1,2$ is the layer index. The quadratic part of the Hubbard Hamiltonian contains electron hopping and a chemical potential $\mu$ terms,
\begin{eqnarray}
	H_K &=& -t \sum_{\langle ij \rangle \ell \sigma} c^\dagger_{i \ell \sigma} c_{j \ell \sigma}
		- t_\perp \sum_{i \sigma} \left( c^\dagger_{i1\sigma} c_{i2 \sigma} + h.c. \right)
	\nonumber \\
	&&- \sum_{i \ell \sigma} \mu_\ell n_{i \ell \sigma}
	\label{BilayerBeginEq}
\end{eqnarray}
where $\langle ij \rangle$ is a sum over in-plane nearest-neighbors, $t$ is the in-plane hopping, and $t_\perp$ the interlayer hopping. We choose the chemical potential in both layers to be opposite, $ \mu_1 = - \mu_2$, in order to induce $p$-doping in one layer and $n$-doping in the other. The Hubbard model additionally contains an on-site electron repulsion,
\begin{equation}
	H_U= U \sum_{i \ell} 
	\left[ ( n_{i \ell \uparrow} - \frac{1}{2} )
		( n_{i \ell \downarrow} - \frac{1}{2} ) 
		- \frac{1}{4} \right]
\end{equation}
which we have written such that $\mu=0$ corresponds to half-filling.

The formation of an exciton condensate requires interlayer electron-hole attraction, which can be incorporated by including a nearest-neighbor interlayer electron-electron repulsion,
\begin{equation}
	H_V = V \sum_{i \sigma \sigma'}
		\left[ ( n_{i 1 \sigma} - \frac{1}{2} )
		( n_{i 2 \sigma'} - \frac{1}{2} ) 
		- \frac{1}{4} \right].
\end{equation}
Note that with the inclusion of $H_V$, $\mu=0$ still corresponds to half-filling. The full Hamiltonian thus consists of the electron hopping, on-site repulsion and an interlayer repulsion,
\begin{equation}
	H = H_K + H_U + H_V.
	\label{HVCh3}
\end{equation}
We study this model using the framework of DQMC, which we briefly summarize here following Refs. \cite{Blankenbecler:1981p3063,White:1989p3778,White:1989p4002}, and emphasizing the changes that are required for the bilayer model with interlayer interactions.

\subsection{The DQMC Algorithm}

In absence of interactions, the kinetic part of the bilayer Hubbard model $H_K$ can be cast into a matrix form,
\begin{equation}
	H_K = \sum_{i j \ell \ell' \sigma}
		c^\dagger_{i \ell \sigma} 
		k_{i\ell,j\ell'}^{\sigma} c_{j \ell' \sigma}.
\end{equation}
We work on a lattice where each layer has $N \equiv N_x \times N_y$ ($N_x=N_y$) sites. Consequently the matrix $k^\sigma$ is a $2N \times 2N$ matrix. In the non-interacting limit, the partition function and the Greens function can now be exactly computed using determinants of this $k$-matrix,
\begin{equation}
	Z \equiv \Tr \left[ e^{-\beta H_K} \right]
	= \det \left[I_{2N} + e^{-\beta k^\uparrow} \right]
		\det \left[I_{2N} + e^{-\beta k^\downarrow} \right]
	\label{QuadrZ}
\end{equation}
and
\begin{equation}
	G^{\sigma}_{i\ell,j\ell'} \equiv \frac{1}{Z}	\Tr \left[ 
		c_{i \ell \sigma} c^\dagger_{j \ell' \sigma}
		 e^{- \beta H_K } \right]
	= \left[ I_{2N} + e^{-\beta k^\sigma} \right]^{-1}_{i\ell, j \ell'}
\end{equation}

To include the interactions $H_U$ and $H_V$ we need a Suzuki-Trotter decomposition, whereby the inverse temperature $\beta$ is considered as a new dimension and the imaginary time axis is cut into $L$ discrete intervals $\Delta \tau = \beta / L$ in length,
\begin{equation}
	e^{-\beta H} \approx \left( e^{-\Delta \tau H_U} e^{-\Delta \tau H_V} e^{-\Delta \tau H_K} \right)^L
\end{equation}
where errors of order $(\Delta \tau)^2$ and higher are neglected. The interactions $H_U$ and $H_V$ can be brought into a quadratic form by a discrete Hubbard-Stratonovich (HS) transformation. On each site and time-slice an Ising variable $s(i,\tau)$ is introduced for each type of interaction,
\begin{equation}
	e^{-\Delta \tau  U \left[ 
		( \hat{n}_{ \uparrow} - \frac{1}{2} )
		( \hat{n}_{ \downarrow} - \frac{1}{2} ) 
		- \frac{1}{4} \right]} 
	= \frac{1}{2} \sum_{s = \pm 1}
		e^{\lambda_U s 
			(\hat{n}_\uparrow - \hat{n}_\downarrow) },
\end{equation}
where $\lambda_{U} = \mathrm{arccosh} \left( e^{\frac{1}{2} U \Delta \tau} \right).$ A similar decoupling works for the interlayer interaction $V$. For each in-plane coordinate $i$ we now have six HS fields $s_\alpha(i,\tau)$: $s_1$ and $s_2$ are associated with the on-site repulsion $U$ in each layer, and the four possible fields $s_3 - s_6$ are associated with the four spin-dependent interlayer interactions $V$.

As the interaction terms have become quadratic, one can rewrite the transformed interaction terms using a diagonal matrix $v^\sigma(\tau)$ that depends on the HS fields $s_\alpha(i,\tau)$. Each time slice is therefore represented by the $2N \times 2N$ matrix
\begin{equation}
	B^\sigma_l = e^{ v^\sigma (l \Delta \tau)} e^{- \Delta \tau k^\sigma},
\end{equation}
the product of which represents the full evolution among the imaginary time axis
\begin{equation}
	M^\sigma = I_{2N} + B^\sigma_L B^\sigma_{L-1} 
		\cdots B^\sigma_2 B^\sigma_1.
\end{equation}
At this level, the partition function can be computed exactly by 
\begin{equation}
	Z = \frac{1}{2^{6NL}} \sum_{s_\alpha (i,\tau)=\pm 1} 
		\det M^\uparrow \det M^\downarrow.
	\label{Zexact}
\end{equation}
Similarly, the Greens function can be computed by
\begin{equation}
	G^\sigma_{i\ell,j\ell'} = \frac{1}{Z} \frac{1}{2^{6NL}} 
		\sum_{s_\alpha(i,\tau)=\pm 1}
		\left[ M^\sigma \right]^{-1}_{i\ell,j\ell'}
		\det M^\uparrow \det M^\downarrow.
	\label{Gexact}
\end{equation}
The bilayer Hubbard model can thus be simulated using standard Ising importance sampling Monte Carlo techniques. The weight of each HS field configuration is given by the absolute value of the determinants
\begin{equation}
	P(s) = \left| \det M^\uparrow \det M^\downarrow \right|.
\end{equation}
such that all measured quantities should be normalized by the average sign of the determinants. When the average sign approaches zero, the statistical errors associated with measurements increases drastically. We will come back to this problem in section \ref{SecFermi}.

\subsection{Sampling the HS fields}

The evaluation of the Greens function Eq. (\ref{Gexact}) for a given configuration of HS fields requires $O(N^3)$ operations. Within the original DQMC approach, this numerical complexity can be reduced by an efficient single-site update scheme.\cite{Blankenbecler:1981p3063} Thereby the flip of the HS field at a single site and time-slice is proposed. If the change is accepted, the new Greens function can be computed by $O(N^2)$ operations. In the bilayer set-up presented here, however, we are dealing with six HS fields per site instead of just one. Therefore, the traditional Sherman-Morrison update cannot be used. We propose a generalization of the aforementioned scheme, based on the Woodbury matrix identity.\cite{Woodbury:1950p5678} 

Starting with a known Greens function at time-slice $l$ for both spin species given the HS-fields $s_\alpha (i,l \Delta\tau)$, we propose random changes $s \rightarrow \pm s$ on all the six HS fields at site $i$ and time-slice $l$ individually. This defines a matrix
\begin{equation}
	\Delta s_\alpha = s_{\alpha,\mathrm{new}} (i,l \Delta\tau) - s_{\alpha,\mathrm{old}} (i,l \Delta\tau) = \pm 2 \; \mathrm{or} \; 0.
\end{equation}
Under this change, we note that the change in the matrix product
\begin{eqnarray}
	A^\sigma & = &
		B^\sigma_l B^\sigma_{l-1} \cdots B^\sigma_1 B^\sigma_{L} \cdots B^\sigma_{l+1}
		\\
	& \rightarrow &
		A^\sigma \, ' = 
		\left[ I + \Delta^\sigma \right] A^\sigma.
\end{eqnarray}
can be factored such that the matrix $\Delta^\sigma$ has only two nonzero elements, namely
\begin{eqnarray}
	\Delta^\uparrow_{i1, i1} & = &
		\exp \left[ \lambda_U \Delta s_1 + \lambda_V ( \Delta s_3 + \Delta s_4) \right] - 1 \\
	\Delta^\downarrow_{i1,i1} & = &
		\exp \left[ - \lambda_U \Delta s_1 + \lambda_V ( \Delta s_5 + \Delta s_6) \right] - 1 \\
	\Delta^\uparrow_{i2, i2} & = &
		\exp \left[ \lambda_U \Delta s_2 - \lambda_V ( \Delta s_3 + \Delta s_5) \right] - 1 \\
	\Delta^\downarrow_{i2,i2} & = &
		\exp \left[ - \lambda_U \Delta s_2 - \lambda_V ( \Delta s_4 + \Delta s_6) \right] - 1.
\end{eqnarray}
Thus the ratio of weights between the new $M^\sigma \, '$ and the old $M^\sigma$ equals
\begin{eqnarray}
	R^\sigma &=& 
	\frac{ \det [ I + A^\sigma \, '  ]}
		{ \det [ I + A^\sigma \,  ] } 
		\nonumber \\
	& = &
		\left[ 1 + (1 - G^\sigma_{i1,i1} ) \Delta^\sigma_{i1,i1} \right]
		\left[ 1 + (1 - G^\sigma_{i2,i2} ) \Delta^\sigma_{i2,i2} \right]
		\nonumber \\ &&
		- G^\sigma_{i2,i1} G^\sigma_{i1,i2} \Delta^\sigma_{i1,i1} \Delta^\sigma_{i2,i2}
	\label{DefRSigma}.
\end{eqnarray}
Using this formula, one can quickly decide whether to accept a change or not
. Under the change of HS fields the new Greens function can be easily computed
\begin{equation}
	G^\sigma \rightarrow G^\sigma \, ' 	= \left[ 1 + A^\sigma 
		+ \Delta^\sigma  A^\sigma  \right]^{-1}.
\end{equation}
Because the matrix $\Delta$ is sparse (it has only two nonzero elements, at $i$ and $i+N$), we can use the Woodbury matrix identity to do a fast update of the Greens function, which is a generalization of the Sherman-Morrison matrix identity. The Woodbury identity amounts to
\begin{equation}
	G_{ab}^\sigma \rightarrow
		G_{ab}^\sigma - \sum_{\ell, \ell'}
			G_{a, i\ell}^\sigma \Delta^\sigma_{i\ell, i\ell} 
			(R^{-1})_{\ell \ell'} (1 - G^\sigma)_{i \ell', b}
\end{equation}
where the spin-dependent $R$-matrix is given by
\begin{equation}
	R_{\ell \ell'} = 
	\begin{pmatrix}
		 1 + (1 - G^\sigma_{i1,i1} ) \Delta^\sigma_{i1,i1} 
		 	&&  - G^\sigma_{i1,i2} \Delta^\sigma_{i2,i2} \\
		 - G^\sigma_{i2,i1} \Delta^\sigma_{i1,i1} 
		 	&& 1 + (1 - G^\sigma_{i2,i2} ) \Delta^\sigma_{i2,i2}
	\end{pmatrix}.
	\label{Rmatrix}
\end{equation}
Note that the determinant of the $R$-matrix equals the ratio as defined in equation (\ref{DefRSigma}). The inverse $(R^{-1})_{\ell \ell'}$ can be computed analytically and  equals
\begin{equation}
	\frac{1}{R^\sigma} \begin{pmatrix}
		1 + (1 - G^\sigma_{i2,i2} ) \Delta^\sigma_{i2,i2}		  
		 	&&  G^\sigma_{i1,i2} \Delta^\sigma_{i2,i2} \\
		G^\sigma_{i2,i1} \Delta^\sigma_{i1,i1} 
		 	&& 1 + (1 - G^\sigma_{i1,i1} ) \Delta^\sigma_{i1,i1}
	\end{pmatrix}.
\end{equation}
These equations combined constitute the full Woodbury update.

To decide whether a proposed change will be accepted we used a linear combination of the Metropolis and the heat bath algorithm. The transition amplitude from one configuration $s$ to the next $s'$ is given by
\begin{equation}
	T(s \rightarrow s') = \gamma \frac{P(s')}{P(s)+P(s')}
		+ (1-\gamma) \min\left( 1,\frac{P(s')}{P(s)} \right).
	\label{TransAmpl}
\end{equation}
The parameter $\gamma$ is tuned self-consistently to achieve an acceptance ratio of about $50 \%$.

Using the Woodbury update scheme presented here one can perform fast updates of all the HS fields at all sites and time-slices. All data points in this paper are obtained by doing $5 \times 10^4$ full sweeps of the lattice for 8 independent Markov chains. Equal-time measurements are averaged over different time-slices, yielding in total $2 \times 10^6$ data points. Unequal-time measurements are performed after every 10 full sweeps, yielding $4 \times 10^5$ data points. The statistical error bars are obtained from the sample variance over the 8 independent chains.

We computed various physical properties, to be discussed in the following sections, for fixed $U=4t$, where $t=1$ serves as a unit of energy. Unless otherwise stated the interlayer hopping is $t_\perp = 0.05t$. We consider various different values of the chemical potential, which is chosen to be opposite in the two layers, $\mu_1 = -\mu_2$. The interlayer interaction $V$ is chosen on a range from $V=0$ to $V=1.25t$. Realistic materials have indeed $V<U$ since Coulomb repulsion is naturally weaker for nearest neighbor than for onsite repulsion. 

All results were obtained on a $N_x =N_y= 4$ and $N_x=N_y=6$ bilayer. Such a bilayer contains $N=2(N_x)^2$ lattice sites; note that the $N_x=6$ bilayer with its 72 sites is numerically more involved than the canonical $N_x=8$ single layer simulations.

\section{Fermion sign problem}
\label{SecFermi}

The average sign limits the applicability of the DQMC method when dealing with strongly correlated electrons. We therefore present first our results regarding the average value of the fermion sign with regards to both its value and its impact on the statistical errors on other measurements.

\begin{figure}
 \includegraphics[width=\columnwidth]{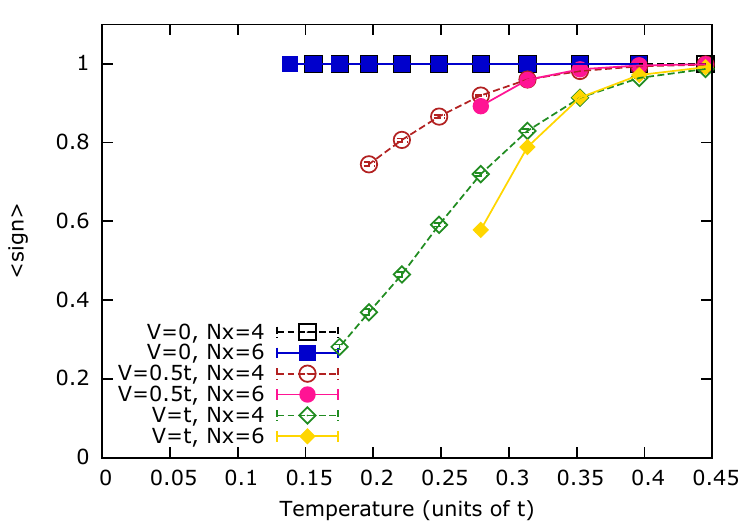}
 \caption{The average sign at half-filling for various interlayer interactions $V$. When $V\neq0$ the average sign drops rapidly. Parameters are $U=4t$, $t_\perp = 0.05t$ and $\mu=0$.}
 \label{FigSignHalffilling}
\end{figure}

In the absence of the interlayer interaction $V$, at half-filling, the average sign is protected by particle-hole symmetry\cite{Hirsch:1985p2854}. This can be understood by considering the determinants of $M^\sigma$ for both spin species. Since the Hubbard-Stratonovich fields couple to both the up and down spins, a change of sign in $\det M^\uparrow$ is accompanied by the same sign change in $\det M^\downarrow$. Consequently, at half-filling for $V=0$ the average sign is always equal to 1. One can directly infer that the sign protection fails when $V \neq 0$, since there are now HS fields that couple to only one type of spin. Indeed, as can be seen in Fig. \ref{FigSignHalffilling}, inclusion of a nonzero interlayer interaction $V$ drastically reduces the sign at half-filling.

\begin{figure}
 \includegraphics[width=\columnwidth]{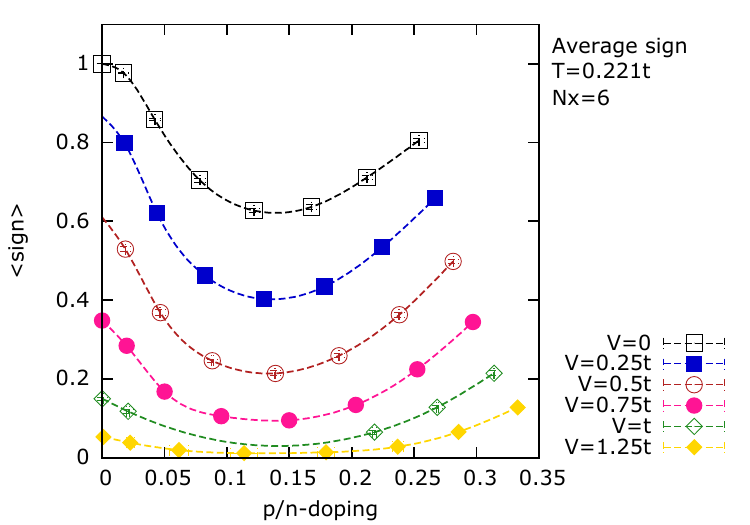}
 \caption{The average sign versus density for increasing interlayer interaction $V$. The temperature is fixed at $T=0.221t$. The remaining parameters are $U=4t$ and $t_\perp = 0.05t$.}
 \label{Fig3-3}
\end{figure}

\begin{figure}
 \includegraphics[width=\columnwidth]{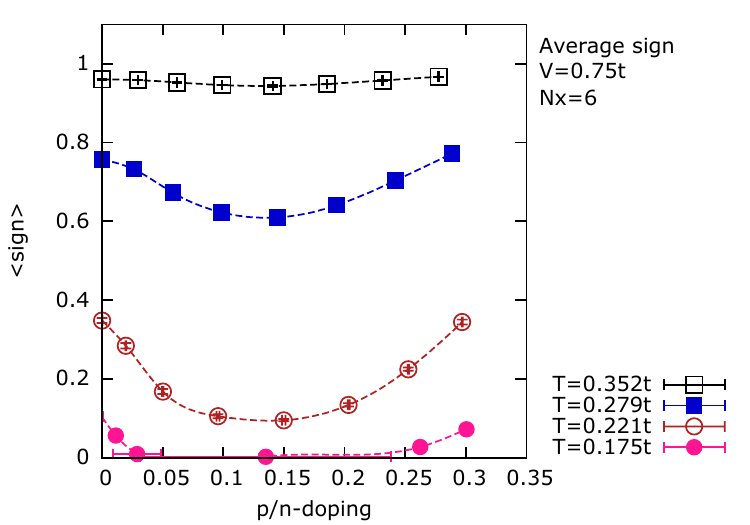}
 \caption{The average sign versus density at various temperatures. The interlayer interaction is fixed at $V=0.75t$, other parameters are $U=4t$ and $t_\perp = 0.05t$. The sign is lowest around $15-20\%$ doping.}
 \label{Fig3-2}
\end{figure}

At finite densities the sign problem becomes increasingly severe. Even in the absence of interlayer $V$, the sign drops so rapidly that the parameter regime physically relevant for the pseudogap, $d$-wave superconductivity, etc., is inaccessible at low temperatures. Fig. \ref{Fig3-3} displays how the average sign depends on both doping and interlayer $V$ for a fixed temperature. It is worthwhile to note that the physical temperature corresponding to these parameters is about 900 Kelvins, still an order of magnitude higher than for example the onset of superconductivity in the cuprates. In Fig. \ref{Fig3-2} we show the average sign as a function of doping for a fixed interlayer coupling $V=0.75t$. 
In all cases the sign problem is the worst around $15-20\%$ doping. This means that in our set-up that we have one layer doped with holes and another layer doped with the same number of electrons, relative to half-filling. Inclusion of $V$ does not change the qualitative doping dependence of the average sign but it does worsen the situation significantly.

Our results for the fermion signs are in contrast to the suggestion, made in the context of the exciton $t-J$ model,\cite{Sheng:1996p5102,Weng:2008p6,Rademaker:2012p5327} for the limit $V \gg t$, that the sign problem could be reduced upon increasing $V$. The increased interlayer interaction makes it more likely that electrons and holes in the two layers move simultaneously, which suggests that the signs of the electrons could be cancelled by the signs of the hole. Our DQMC results at indicate that this is clearly not the case when the interlayer coupling $V$ is of the same order as $t$. However, it is still a possibility that the average sign increases when a condensate forms.

Whenever the average sign becomes small, the uncertainty of the measurements increases. Let us construct a rough quantitative estimate of the floor in the value of the average sign at which simulations are still feasible. For this purpose we need to distinguish between two kinds of measurements. The first are `static' measurements, such as the doping and interlayer tunneling, which involve only the equal-time Greens function
\begin{equation}
	G^{\sigma}_{i\ell,j\ell'} = \langle c_{i \ell \sigma} c^\dagger_{j \ell' \sigma} \rangle
\end{equation}
The second are `dynamic' measurements involving the unequal-time Greens function
\begin{equation}
	G^{\sigma}_{i\ell,j\ell'} (\tau) = \langle T_\tau c_{i \ell \sigma} (\tau) c^\dagger_{j \ell' \sigma} \rangle
\end{equation}
which are generally less stable than static measurements. As an example of the latter category, we have computed the dc conductivity from the current-current correlation function for each layer,
\begin{equation}
	\Lambda_{xx}^\ell (\mathbf{q}, \tau)
		= \sum_{i} \langle j_x (\mathbf{r}_{i\ell}, \tau) j_x (0,0) \rangle e^{i \mathbf{q} \cdot \mathbf{r}_{i\ell}}
	\label{dc1}
\end{equation}
Instead of performing the analytic continuation, we approximate the dc conductivity by\cite{Trivedi:1996p3998}
\begin{equation}
	\frac{ \pi \sigma_{dc}}{\beta^2} = \Lambda_{xx}^\ell (\mathbf{q}=0, \tau=\beta/2).
	\label{dc2}
\end{equation}

\begin{figure}
 \includegraphics[width=\columnwidth]{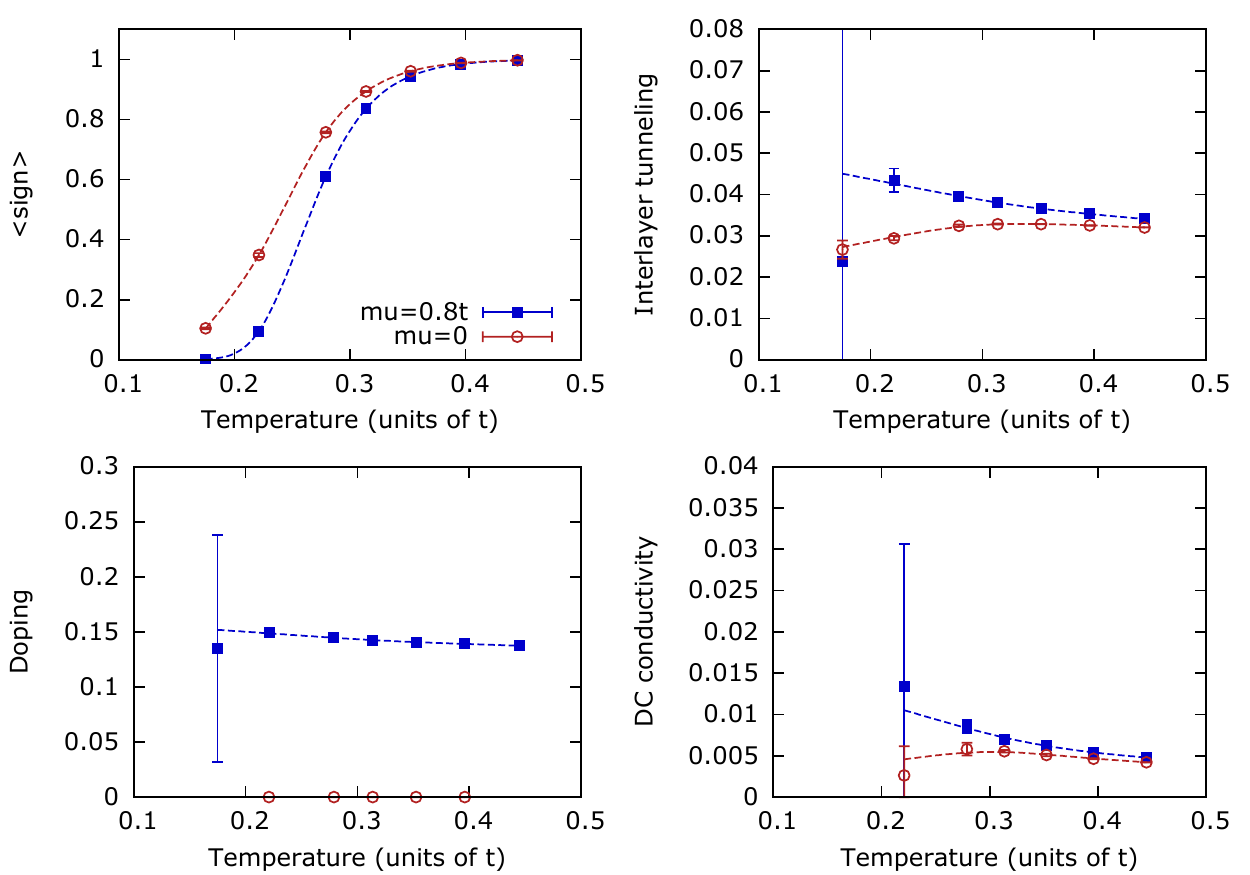}
 \caption{Average sign, doping, interlayer tunneling and dc conductivity for $V=0.75t$ and $N_x=6$ as a function of temperature. Static measurements, such as density and interlayer tunneling, are still reliable as long as the sign $> 0.05$. The dynamic measurements such as dc conductivity become unreliable when the sign $<0.5$. For comparison, both $\mu=0$ and $\mu=0.8t$ is shown.}
 \label{Fig3-5}
\end{figure}

In Fig. \ref{Fig3-5} we show the average sign as function of temperature for $V=0.75t$, for $\mu=0$ and $\mu=0.8t$. In addition, we also show the measurements of the interlayer tunneling, doping and dc conductivity. As long as the average sign is above $0.5$, all measurements are statistically trustworthy. Below $0.5$, the results for the dc conductivity have statistical error bars more than half of $\sigma_{dc}$ itself. Therefore we limit our dynamical measurements to regions where the sign is $> 0.5$. Similarly, as long as the sign is greater than $\sim 0.1$ the statistical error on static measurements is manageable. This implies that the window for which DQMC is applicable for all doping levels is limited to about $\beta < 5/t$ and $V < t$.

\section{Exciton condensation}
\label{SecEC}
Our main goal is to investigate whether exciton condensation might occur in the bilayer Hubbard model. To do so, we examine the order parameter of an interlayer exciton condensate, which is defined as
\begin{equation}
	\Delta_{\mathbf{k}}
	= \langle c^\dagger_{\mathbf{k}1 \sigma} c_{\mathbf{k} 2 \sigma} \rangle.
	\label{OrderParameter}
\end{equation}
In the presence of strong local interactions excitons will be formed locally, such that the electron and hole are on the same interlayer rung. The order parameter becomes independent of momentum and equals
\begin{equation}
	\Delta
	= \frac{1}{N} \sum_i \langle c^\dagger_{i 1 \sigma} c_{i 2 \sigma} \rangle.
	\label{OrderParameter2}
\end{equation}
Consequently, the condensate order parameter equals the expectation value of the interlayer tunneling,\cite{Spielman:2000p4834,Eisenstein:2004p4770} which is directly measurable in experimental setups. 

Within the DQMC method, the interlayer tunneling follows directly from the Greens function. The ideal exciton condensate occurs when the interlayer hopping is completely suppressed, $t_\perp=0$. However, in that case the order parameter calculated in DQMC is identically zero. We need to include a finite $t_\perp$, which acts as a symmetry breaking field just as a magnetic field induces magnetization. We thus construct a way to approach the limit $t_\perp \rightarrow 0$ in order to decide whether there are signs of spontaneous symmetry breaking. Let us consider two approaches.

\begin{figure}
 \includegraphics[width=\columnwidth]{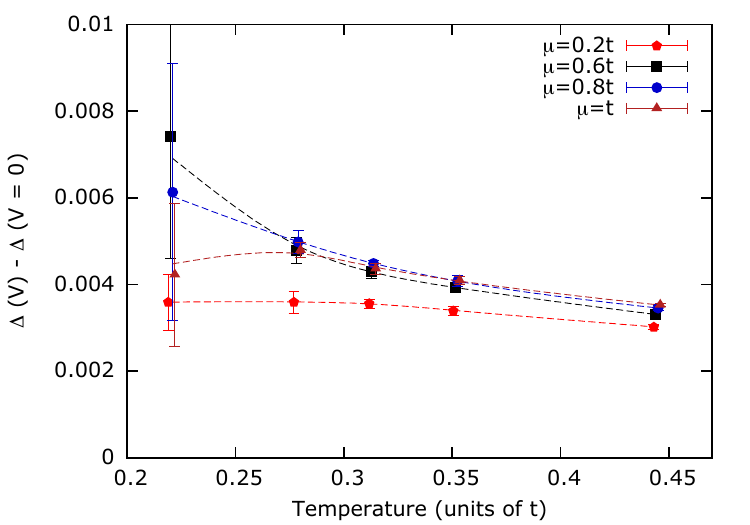}
 \caption{Interlayer tunneling at $V=0.75t$ for $N_x=6$, relative to the $V=0$ case. A clear enhancement of the tunneling amplitude, which is equal to the order parameter of the exciton condensate, can be seen around $\mu=0.8t$, where the doping level is approximately $15\%$.}
 \label{Fig4-1}
\end{figure}
\begin{figure}
 \includegraphics[width=\columnwidth]{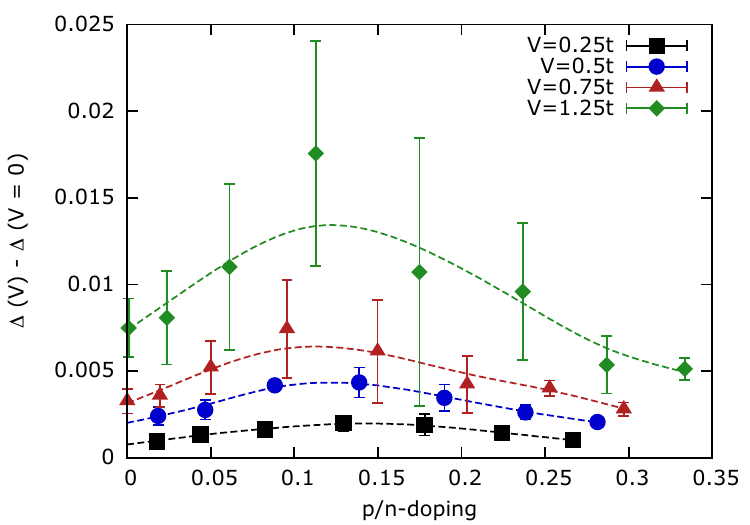}
 \caption{Interlayer tunneling for $N_x=6$ at $T=0.221t$, relative to the $V=0$ case, for all densities and interaction $V$. A clear enhancement of the tunneling, which is equal to the exciton condensate order parameter, can be seen around the doping level of $10-20\%$.}
 \label{Fig4-6}
\end{figure}

A first approach follows from the notion that in the absence of an exciton pairing interaction $V$ the interlayer hopping $t_\perp$ will automatically create a doping dependence of the interlayer tunneling. Therefore we separate the contribution to interlayer tunneling that arises due to exciton formation from the part that is already present at $V=0$. In Fig. \ref{Fig4-1} we show how this relative interlayer tunneling depends on temperature and chemical potential for fixed $V=0.75t$. There is a clear enhancement of the interlayer tunneling around $\mu=0.8t$, which amounts to $15 \%$ doping. At a given temperature of $T=0.221t$ we present the interlayer tunneling as a function of doping and interlayer interaction $V$ in Fig. \ref{Fig4-6}. The strongest tendency towards interlayer tunneling is indeed at $10-20\%$ doping, for the largest values of interaction $V$. Indeed, once again the most interesting physics seems to happen where the sign problem is most severe.

\begin{figure}
 \includegraphics[width=\columnwidth]{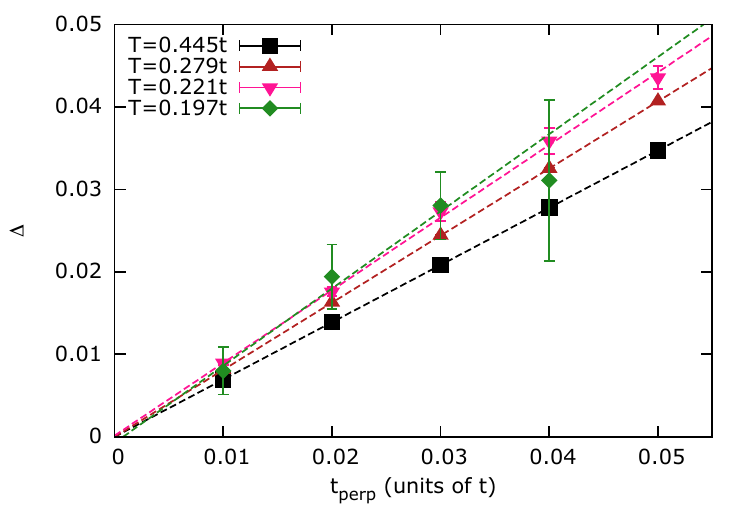}
 \caption{Interlayer tunneling at $V=0.75t$ as a function of $t_\perp$ for $\mu=t$ and $N_x=6$. The scaling for $t_\perp$ suggests that there is no exciton condensation.}
 \label{Fig4-3}
\end{figure}

Our second approach to determine the possibility of exciton condensation is to look at the $t_\perp$-dependence of the interlayer tunneling. Following the standard BEC/BCS condensation theories, the exciton condensate is represented in the Hamiltonian by the symmetry breaking term
\begin{equation}
- V \Delta \sum_{i \sigma} 
	\left( c^\dagger_{i 1 \sigma} c_{i 2 \sigma} 
			+ h.c. \right)
\end{equation}
which adds to the interlayer hopping term $t_\perp$. When $U=0$ and $V$ infinitesimally small we can compute the ground state expectation value of the interlayer tunneling given this order parameter $\Delta$ which yields
\begin{equation}
	\langle c^\dagger_{i 1 \sigma} c_{i 2 \sigma} \rangle
		\sim \frac{t_\perp + V \Delta}{t}.
\end{equation}
For finite $U$ and $V$ we therefore assume that the interlayer tunneling is a linear function of $t_\perp$, and the order parameter can be found by taking the limit $t_\perp \rightarrow 0$. This is done for $V=0.75t$ and $\mu=t$, parameters for which the interlayer tunneling is the largest, in Fig. \ref{Fig4-3}. As the temperature is lowered the interlayer tunneling increases. However, the scaling behavior as a function of $t_\perp$ suggests that there is no exciton condensation present at the temperatures considered. Due to the sign problem, we cannot reliably access lower temperatures.

\begin{figure}
 \includegraphics[width=\columnwidth]{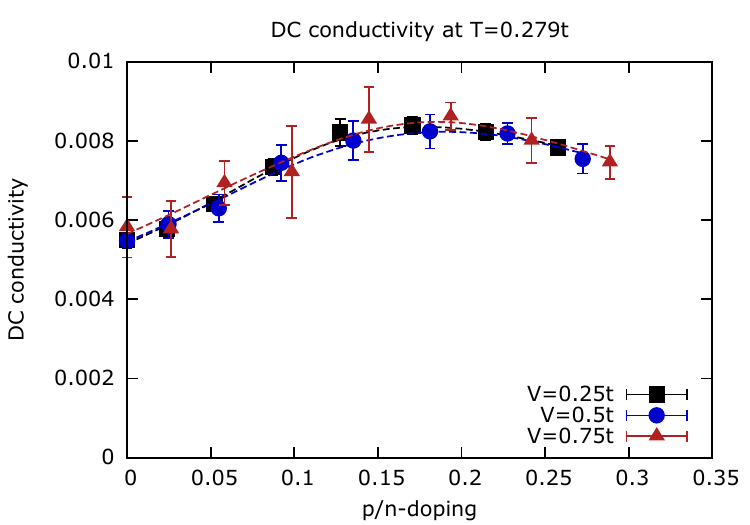}
 \caption{The dc conductivity $\sigma_{dc}$ following equations (\ref{dc1})-(\ref{dc2}) at $T=0.279t$ as a function of doping and $V$ for $N_x=6$. The dc conductivity is the largest at a doping around $15-20\%$.}
 \label{Fig4-4}
\end{figure}
\begin{figure}
 \includegraphics[width=\columnwidth]{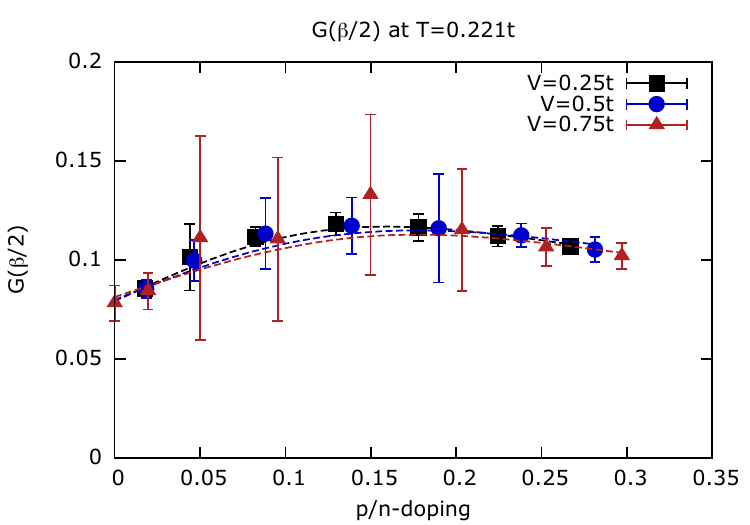}
 \caption{The density of states at the Fermi level, approximated by $G(\beta/2)$, at $T=0.221t$ for $N_x=4$. The density of states is highest around 15-20\% doping, independent of the interlayer interaction $V$.}
 \label{Fig4-5}
\end{figure}

Next we turn to direct measurements of the order parameter and experimental probes related to properties of the exciton condensate. Since in an exciton condensate the charge carriers are bound into charge neutral excitons, it is expected that exciton condensates are insulating. We therefore perform conductivity measurements. In Fig. \ref{Fig4-4} we display measurements on the dc conductivity following Eqs. (\ref{dc1}) and (\ref{dc2}). The conductivity is largest at a doping of 15-20\% and is independent of $V$, indicating more metallic behavior. Alternatively, one can look at the density of states at the Fermi level, which is approximated by $G(\mathbf{r}=0,\tau=\beta/2)$.\cite{Trivedi:1995p5680,Nowadnick:2012p5679} As expected, the density of states at the Fermi level follows the same trend as the dc conductivity. We therefore conclude that we have found no evidence of exciton condensation in the bilayer Hubbard model in the parameter and temperature regime accessible by DQMC. However, the increased interlayer tunneling suggests that exciton physics might be relevant for large $V$ and around 10-20\% doping.

\section{Magnetic measurements}
\label{SecMag}
Strong correlations can lead to the localization of electron degrees of freedom, resulting in magnetic correlations. For the Hubbard model on a square lattice this results in antiferromagnetic order at half-filling. Experiments on the cuprates, however, suggest that this antiferromagnetism quickly disappears upon doping.\cite{Imada:1998p2790} In this section we will therefore study the influence of both doping and interlayer interactions on the antiferromagnetic state.

\begin{figure}
 \includegraphics[width=\columnwidth]{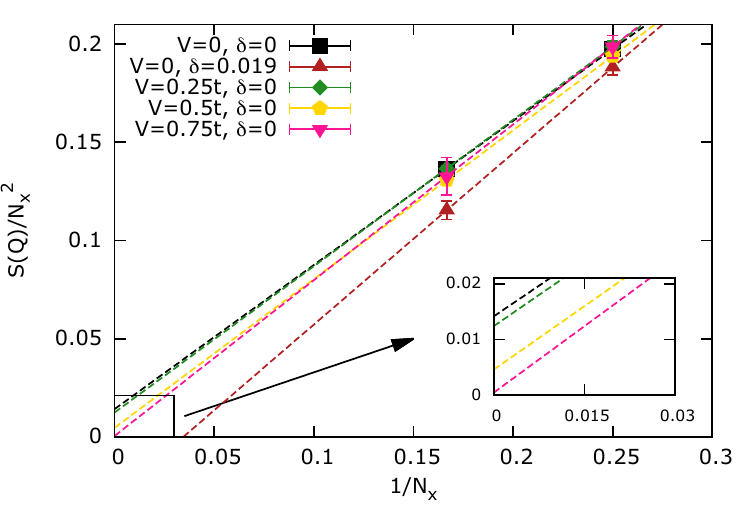}
 \caption{Antiferromagnetic correlations at $T=0.175t$ for various $V$ and doping. Only at half-filling ($\delta=0$) we find antiferromagnetism in the thermodynamic limit. The antiferromagnetism remains when a nonzero $V$ is included.}
 \label{Fig5-1}
\end{figure}

The antiferromagnetic structure factor in each layer is given by
\begin{equation}
	S_{\ell} (\mathbf{Q}) = \frac{1}{N_x^2} \sum_{i j} e^{i \mathbf{Q} \cdot (\mathbf{r}_i - \mathbf{r}_j)}
		\langle \left(n_{i \ell \uparrow} - n_{i \ell \downarrow} \right)
		\left(n_{j \ell \uparrow} - n_{j \ell \downarrow} \right) \rangle
\end{equation}
where $\mathbf{Q} = (\pi/a,\pi/a)$ is the antiferromagnetic wave vector and $a=1$ is the in-plane lattice constant. Spin wave theory\cite{Huse:1988p5681} suggests that $S(\mathbf{Q})$ scales with $1/N_x$ on a finite cluster. The thermodynamic limit $N_x \rightarrow$ of $S(\mathbf{Q})$ can be found from a linear extrapolation of the $N_x=4$ and $N_x = 6$ data, as is done in figure \ref{Fig5-1}. Indeed, the antiferromagnetic order is rapidly destroyed as one dopes the layers. However, under the inclusion of $V$ the antiferromagnetic order remains up to $V=0.75t$. 

\begin{figure}
 \includegraphics[width=\columnwidth]{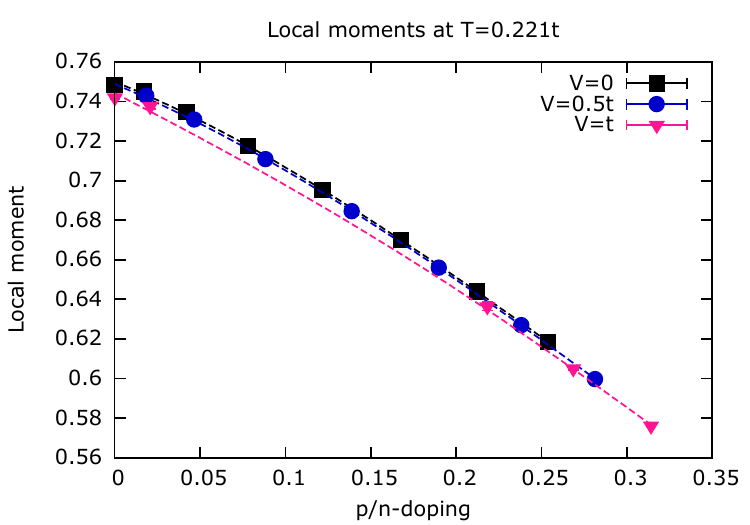}
 \caption{The local moments as a function of $V$ and doping at $T=0.221t$ for $N_x=6$. The localization of electrons is the strongest at half-filling, and almost independent of the interlayer interaction $V$.}
 \label{Fig5-2}
\end{figure}

Even though the antiferromagnetic order is rapidly destroyed, the localization of electrons associated with the strong onsite repulsion $U$ is barely reduced. The local moment, which measures the degree of localization, is defined as
\begin{equation}
	m_{i \ell}
	 = \langle \left(n_{i\ell \uparrow} - n_{i \ell \downarrow} \right) \rangle.
\end{equation}
The site-averaged local moments are shown in Fig. \ref{Fig5-2}. The localization of electrons is rather independent of the interlayer interaction strength $V$.

\section{Conclusions}

In conclusion, we have performed a DQMC simulation of the extended bilayer Hubbard model to study possible exciton condensation in $p$/$n$-doped bilayers. With the Woodbury update we introduced a method that allows DQMC studies of systems with multiple HS fields. Because the different HS fields are treated on an equal level, this allows for numerical studies of interacting systems with competing orders. The expectation that with the formation of bosonic excitons the fermionic sign problem would be reduced\cite{Bouadim:2008p3331} cannot be confirmed within this approach, though Monte Carlo simulations of the strongly coupled $t$-$J$ model may have less severe sign problems\cite{Rademaker:2013p5631}. Given the average value of the fermion signs we measured, we conclude that the applicability of the DQMC is limited to about $\beta < 5/t$ and $V < t$. In this regime there is no direct indication of exciton condensation, but the interlayer tunneling results indicate a strong exciton tendency around $10-20 \%$ $p$/$n$-doping, which can be interpreted as a precursor to condensation. The question thus remains whether at low enough temperatures exciton condensation is possible in correlated bilayers. It is interesting to note, however, that several properties, such as the local moments, magnetic order, conductivity and the density of states, seem to be independent of the interlayer coupling $V$. Why such interlayer interaction $V$ apparently does not influence these properties is another open and interesting problem.

\acknowledgments{This research was supported by the Dutch NWO foundation through a VICI grant. The authors thank Kai Wu, Tom Devereaux, Brian Moritz and Hans Hilgenkamp for helpful discussions, and Jeroen Huijben for designing Figure \ref{BilayerHubbardFig}.}


\end{document}